\documentclass[aoas,MSNbibl,nameyear,rotating,seceqn,dvips]{arximspdf}
\usepackage{dcolumn}
\usepackage{graphicx}

\doi{10.1214/13-AOAS634} 
\volume{7}
\issue{3}
\pubyear{2013}
\firstpage{1640}
\lastpage{1662}

\makeatletter
\newcolumntype{d}[1]{D{.}{.}{#1}}
\newtheorem{theorem}{Theorem}
\def\cal{\mathcal}
\newcommand{\ROC}{\operatorname{ROC}}
\newcommand{\logit}{\operatorname{logit}}

\makeatother

\begin{document}
\begin{frontmatter}

\title{Logistic regression analysis with standardized markers\thanksref{T1}}
\runtitle{Logistic regression analysis with standardized markers}

\thankstext{T1}{Supported by NIH Grants U01 CA086368-06,
RO1 GM054438, PO1 CA053996, U24CA086368, and P30CA015704.}

\begin{aug}
\author[a]{\fnms{Ying} \snm{Huang}\corref{}\ead[label=e1]{yhuang@fhcrc.org}},
\author[a]{\fnms{Margaret S.} \snm{Pepe}}
\and
\author[a]{\fnms{Ziding} \snm{Feng}}
\runauthor{Y. Huang, M. S. Pepe and Z. Feng}
\affiliation{Fred Hutchinson Cancer Research Center Public Health
Sciences Division and University of Washington}
\address[a]{Fred Hutchinson Cancer Research Center\\
Public Health Sciences Division\\
1100 Fairview Avenue N.\\
Seattle, Washington 98109\\
USA\\
and\\
Department of Biostatistics\\
University of Washington\\
Seattle, Washington 98195\\
USA\\
\printead{e1}} 
\end{aug}

\received{\smonth{6} \syear{2011}}
\revised{\smonth{1} \syear{2013}}

%
\begin{abstract}
Two different approaches to analysis of data from diagnostic biomarker
studies are commonly employed. Logistic regression is used to fit
models for probability of disease given marker values, while ROC curves
and risk distributions are used to evaluate classification performance.
In this paper we present a method that simultaneously accomplishes both
tasks. The key step is to standardize markers relative to the
nondiseased population before including them in the logistic regression
model. Among the advantages of this method are the following: (i)
ensuring that results from regression and performance assessments are
consistent with each other; (ii) allowing covariate adjustment and
covariate effects on ROC curves to be handled in a familiar way, and
(iii) providing a mechanism to incorporate important assumptions about
structure in the ROC curve into the fitted risk model. We develop the
method in detail for the problem of combining biomarker data sets
derived from multiple studies, populations or biomarker measurement
platforms, when ROC curves are similar across data sources. The methods
are applicable to both cohort and case--control sampling designs. The
data set motivating this application concerns Prostate Cancer Antigen 3
(PCA3) for diagnosis of prostate cancer in patients with or without
previous negative biopsy where the ROC curves for PCA3 are found to be
the same in the two populations. The estimated constrained maximum
likelihood and empirical likelihood estimators are derived. The
estimators are compared in simulation studies and the methods are
illustrated with the PCA3 data set.
\end{abstract}

%

\end{frontmatter}

\section{Introduction}\label{sec1}

As myriads of biomarkers are becoming available from research
laboratories, the demand for more sophisticated statistical analysis
methods increases. For example, an emerging request is to combine
information from multiple sources in evaluating a biomarker's
performance. In addition, biomarkers must be evaluated from multiple
points of view, including, for example, their roles as risk factors and
predictors as well as their classification performance.

Logistic regression analysis has been a mainstay of biostatistical
methodology for evaluating risk factors, particularly in epidemiology
and in therapeutic research. For evaluating biomarker performance,
however, other methods are more appropriate, such as those based on
receiver operating characteristic (ROC) curves [\citet{Pepe2003}] and risk distributions.
Methods that evaluate categorized risk distributions are gaining
popularity and are often called risk reclassification methods. However,
in general, methods for evaluating performance are far less well
developed and have more limited availability than logistic regression
methodology.

In this paper we show that evaluation of biomarker performance can be
achieved within the logistic regression framework if as a preliminary
step one standardizes the marker using the control population to define
the reference distribution for standardization. Previously, in a simple
setting with the biomarker as the only predictor of interest, \citet
{GuAndPepe2010} applied a logistic regression model to standardized
marker values as a rank-invariant approach to estimating the variance
of the empirical ROC curve in sample size calculations. Here we extend
the approach to estimate the ROC curve itself and to allow for
additional covariates. Through recognition of the fact that there is a
direct functional relationship between the coefficient for the
standardized marker in the logistic regression model and the ROC curve,
we show that risk distributions as well as the ROC curve conditional on
covariates can be calculated directly from coefficients in the model
and that incorporating this relationship into estimation leads to
efficiency gains. Since the method simultaneously evaluates the marker
as a risk factor and the marker's performance as a classifier, it
provides a more coherent approach than current methods that separately
evaluate the two aspects. Potentially inconsistent results are avoided.
The framework has several other attributes. First, estimated ROC curves
can be constrained to be concave if desired. Concavity is a fundamental
property of the ROC curve that is not taken advantage of by most
standard methods. Second, covariate effects on biomarker performance
can be addressed naturally within the logistic regression framework.
Third, the method is rank invariant with respect to the marker, adding
a degree of robustness compared with usual logistic regression.

The general framework and methods for estimation are presented in
Section~\ref{sec2}. The method is then developed in some detail for the problem
of evaluating a biomarker using data derived from multiple studies or
populations when a common ROC curve across data sources is of interest.
This is an important problem for which methodology has not been
proposed heretofore. Consider that prior to launching a large
validation study, data from multiple small studies, possibly using
different assay platforms, may be examined. Methods that can combine
information across studies are needed in this setting. Moreover, when a
large collaborative validation study is undertaken, it typically
involves multiple sites that may each follow somewhat different
protocols or involve multiple subpopulations that differ in regard to
patient characteristics. Again, in this setting, our methods for
combining data based on a common ROC curve will be useful. The example
that motivated our work concerns evaluating a prostate cancer biomarker
in two subpopulations of men.

In Section \ref{sec3} we describe this example in detail and develop three
methods for estimating parameters in the logistic regression model. We
evaluate the properties of these estimators in simulation studies and
describe results in Section \ref{sec4}. In Section \ref{sec5} we illustrate the
methodology using the data set from the prostate cancer biomarker study
and finish with some concluding remarks in Section \ref{s:discuss}.

\section{The general framework: Logistic regression applied to
standardized biomarker values}\label{sec2}

\subsection{The risk model is related to the ROC curve}\label{sec2.1}

Consider a binary outcome~$D$, disease, say, with $D=0$ for control
nondiseased subjects and $D=1$ for case diseased subjects, a single
continuous marker $Y$, and additional covariates denoted by $X$. Note
the marker $Y$ may be a predefined combination of predictors. For
example, the Framingham risk score is a linear combination of risk
factors for cardiovascular events including age, total cholesterol, and
systolic blood pressure. Another example is the Oncotype-DX recurrence
score that is a fixed combination of 21 gene expression assays. For an
observation with marker measurement $Y=y$ and covariate value $X=x$,
let the standardized marker value be $U=U(x,y)=P(Y>y|D=0,X=x)$. That
is, using as a reference the marker distribution among controls with
covariate value $x$, $U$ is the proportion of marker values in the
reference distribution exceeding $y$. $U$ has been called the placement
value for $Y$ [\citet{PepeAndCai2004}] and $100 \times(1-U)$ is
recognized as the percentile of $Y$ in the reference population [\citet
{HanleyAndHajian1997,HuangAndPepe2009a}]. Percentiles are commonly
used to standardize growth and lung function measurements for children
and to standardize many laboratory measures. Note that the risk
function can be written in terms of $Y$ or $U$,
$P(D=1|Y,X)=P(D=1|U,X)$. Using the latter formulation, we propose the
following logistic regression model:
%
\begin{equation}
\logit P(D=1|U,X)=\logit P(D=1|X)+ G(U,X,\beta), \label{eqnRiskZ}
\end{equation}
where $G(U,X,\beta)$ is some parametric function of $U$ and $X$
parameterized by $\beta$, and $\logit P(D=1|X)$ is an offset term that
will be entered into the model before the application of the logistic
regression.

Interestingly, we can show that $G(U,X,\beta)$ in this framework
corresponds to the log-derivative of $\ROC_X$, where $\ROC_x$ is the
ROC curve for $Y$ in the covariate specific population with $X=x$.

\begin{theorem}\label{th1}
Under model (\ref{eqnRiskZ}), $\ROC
_x(t)=\int_0^t\exp\{G(u,x,\beta)\}\,du$ and $G(u,x,\beta)=\log\ROC_x^{\prime
}(u)$.
\end{theorem}

A proof of Theorem \ref{th1} can be found in Appendix A of the supplementary
material [\citet{HuangEtal2013}]. The proposed framework (\ref
{eqnRiskZ}) therefore naturally links risk modeling with ROC analysis.
These two analytic tasks are typically undertaken separately in current
practice, leading to disjointed and possibly inconsistent results, as
illustrated in a simulated example in Appendix B of the supplementary
material [\citet{HuangEtal2013}]. Having a unified framework for risk
modeling and for ROC analysis offers a more coherent approach to
biomarker evaluation.

\subsection{Ensuring concave ROC curves}\label{sec2.2}
An important attribute of this framework is that the ROC curve using
the marker as the decision variable can be easily constrained to be
concave if desired. Concavity is a fundamental characteristic of proper
ROC curves [\citet{Egan1975,DorfmanEtal1996}]. In settings where it is
reasonable to assume a monotone relationship between the risk of
disease and the marker, the corresponding ROC curve for the marker is
necessarily concave. Yet concavity has not been strictly enforced in
ROC modeling. Indeed, the classic binormal ROC model that is widely
used in radiology is not constrained to be concave. The binormal ROC
curve assumes the existence of a common monotone transformation that
transforms the marker distributions of cases and controls to normality,
but if the variances of those normal distributions differ, the ROC
curve is not concave. There have been several concave ROC models
proposed in the literature, including the ``bigamma'' ROC curve [\citet
{DorfmanEtal1996}], the ``bilomax'' ROC curve [\citet
{CampbellAndRatnaparkhi1993}], and the ``proper'' binormal ROC curve
[\citet{MetzAndPan1999}]. Use of these models has been limited, however,
partly due to difficulties in their implementation. In our framework
(\ref{eqnRiskZ}), we model $G(u,x,\beta)$, the log-derivative of the
ROC curve directly rather than modeling the ROC curve itself.
Consequently, it is easy to constrain the ROC curve to be concave by
modeling $G(u,x,\beta)$ as a monotone decreasing function of $u$.

\subsection{Incorporating covariates}\label{sec2.3}
Equation (\ref{eqnRiskZ}) can be recognized as showing the association
between the pre-test and post-test risk of disease where $G(U,X,\beta)$
is known as the covariate specific diagnostic likelihood ratio. Using
Bayesian terminology,\vadjust{\goodbreak} $G(U,X,\beta)$ is the Bayes factor that relates
the prior and posterior probabilities of disease. However, very little
methodology exists that exploits its relationship with the ROC curve,
$G(U,X,\beta)=\log\ROC_X^{\prime}(U)$. To our knowledge, the only
previous use of this framework is by \citet{GuAndPepe2010}, who
exploited a simplified version of (\ref{eqnRiskZ}) without covariates for a
different problem of estimating the variance of the empirical ROC curve
in sample size calculation.

The framework provides a mechanism to incorporate covariates $X$ into
the model for the ROC curve through $G(U,X,\beta)=\log\ROC_X^{\prime
}(U)$. This model provides opportunities for evaluating effects of
covariates on the discriminatory power of the marker. To illustrate,
consider the following toy example. Suppose $X$ is comprised of two
subsets $X_1$ and $X_2$ that can each be multivariate, for example,
$X_1=\mathrm{age}$ and $X_2=\mathrm{gender}$. Suppose we are interested in testing
whether $X_2$ affects the marker's classification accuracy. We model
$G(U,X,\beta)$ as $\beta_0+\beta_1r(U)+\beta_2^TX_1+\beta
_3^TX_1r(U)+\beta_4^TX_2+\beta_5^TX_2r(U)$, where $r(u)$ is some
pre-specified function of $u$. Then the null hypothesis $H_0$, that
$X_2$ does not affect the ROC curve, corresponds to zero coefficients
for terms involving $X_2$, that is, $H_0\dvtx \beta_4=\beta_5=0$. The
approach applies in general: we can test whether a subset of covariates
affect the marker's discriminatory accuracy by testing the
corresponding coefficients in the risk model. One appealing property of
(\ref{eqnRiskZ}) is that it naturally separates covariate effects for
marker standardization from covariate effects for ROC analysis.
Specifically, covariates used in deriving $U$ are those affecting the
marker distribution in the control population, whereas covariates
involved in $G(U,X,\beta)$ are those affecting the discriminatory power
of the marker, as characterized by the ROC curve.

Two natural competing methods existing in the literature for covariate
adjustment in ROC estimation using standardized marker values are the
nonparametric ROC adjustment method proposed by Janes and Pepe
(\citeyear
{JanesAndPepe2008,JanesAndPepe2009}) and various semiparametric
regression methods based on the binormal ROC model, such as the ROC-GLM
method [\citet{AlonzoAndPepe2002}] and the pseudo-likelihood method
[\citet{PepeAndCai2004}]. Compared to the nonparametric method, applying
a logistic regression form provides a much more efficient alternative.
As mentioned in Section \ref{sec2.2}, existing semiparametric methods relying on
the binormal ROC curve are not natural for modeling a concave ROC
curve. In addition, as detailed in Sections~\ref{sec3} and~\ref{sec4}, they make
inference using standardized marker values among cases only, whereas
the proposed method utilizes standardized marker values for all
subjects and gains efficiency as a consequence.

\subsection{Estimation}\label{sec2.4}
Consider first a cohort or cross-sectional study. Before fitting the
logistic regression model to estimate regression coefficients, we first
need to estimate the offset term $\logit P(D=1|X)$\vspace*{1pt} and the standardized
marker values $U$. Denote the estimators by $\hat{P}(D=1|X)$ and $\hat
{U}$, respectively. For a cohort or cross-sectional study, $\hat
{P}(D=1|X)$ can be derived in standard fashion using logistic
regression techniques or nonparametrically if $X$ is discrete.
Computation of $\hat{U}$ requires estimating the distribution of $Y$ in
the control population conditional on $X$. Methods have been described
and we refer to \citet{HuangAndPepe2009a} for details. In particular,
they suggest nonparametric methods for discrete $X$, and parametric and
semiparametric methods for continuous $X$. After obtaining $\hat
{P}(D=1|X)$ and~$\hat{U}$, we substitute them into the logistic
regression model (\ref{eqnRiskZ}) to estimate coefficients for $X$ and
$\hat{U}$ using standard logistic regression fitting procedures
including $\logit\hat{P}(D=1|X)$ as an offset term.

Consider a case--control sample where cases and controls are randomly
sampled from the case and control subpopulations. Estimation of $U$ can
be performed as in a cohort or cross-sectional study. Let $\mathit{Sampled}$
denote being sampled into the case--control set. According to Bayes'
theorem, we have
\begin{eqnarray*}
&&\frac{P(D=1|U,X)}{P(D=0|U,X}\frac{P(D=0|X)}{P(D=1|X)}\\
&&\qquad=\frac
{P(D=1|U,X,\mathrm{Sampled})}{P(D=0|U,X,\mathrm{Sampled})}\frac
{P(D=0|X,\mathrm{Sampled})}{P(D=1|X,\mathrm{Sampled})},
\end{eqnarray*}
which implies that equation (\ref{eqnRiskZ}) can be written as
\begin{eqnarray*}
&&\logit P(D=1|U,X, \mathrm{Sampled})\\
&&\qquad=\logit P(D=1|X, \mathrm{Sampled})+ G(U,X,\beta).
\label{eqnRiskCCZ}
\end{eqnarray*}
Thus, the estimate of regression coefficients for (\ref{eqnRiskZ}) can
be obtained by applying an ordinary logistic regression to $D$, $X$,
and $\hat{U}$ to the case--control sample with $\logit\hat
{P}(D=1|X,\mathrm{Sampled})$ entered as an offset, where $\hat
{P}(D=1|X,\mathrm{Sampled})$ is an estimate of $P(D=1|X,\mathrm{Sampled})$ based on the
case--control sample.

After the model coefficients for (\ref{eqnRiskZ}) are obtained either
based on a cohort, a~cross-sectional, or a case--control sample, disease
risk for each subject can be computed by entering $\hat{P}(D=1|X)$ and
the individual's $\hat{U}$ into the estimated model. For a case--control
sample, in order to estimate $P(D=1|X)$, information about prevalence,
$P(D=1)$, has to be acquired externally, for example, from the
literature, from another independent cohort, or from the parent cohort
within which the case--control sample is nested. For evaluation of a
biomarker's classification and prediction performance conditional on
covariates, methods have been developed previously when the risk of
disease conditional on the biomarker and covariates follows a logistic
regression model [\citet{HuangEtal2007,HuangAndPepe2010a}]. These
methods can be applied here, replacing the biomarker value on the
original scale $Y$ with their estimated standardized value $\hat{U}$.
Note that the $\hat{U}$'s are correlated with each other due to use of
the control sample for estimating the reference distributions for
standardization. Hence, standard variance formulae from logistic
regression results do not apply, as they assume independence between
observations. Additional complexity is also introduced by using an
estimated offset term. Instead we propose bootstrap resampling for
estimating variances of parameter estimators and constructing
confidence intervals using percentiles of the bootstrap distribution.
The resampling procedures need to mimic the design of the original
study. Specifically, separate resampling of cases and controls would be
warranted if the study is of case--control design, while for a cohort
study resampling would be at random from the cohort without regard to
outcome status.

\section{Methodology for combining data sources through a common ROC curve}\label{sec3}
\subsection{Combining data sets with common ROC curves}\label{sec3.1}

In the remainder of this manuscript, we focus on a specific problem
where we evaluate biomarker performance combining data from multiple
sources through estimation of a common ROC curve across data sources.
The covariate $X$ characterizes the data source and we show how to use
the logistic regression methodology to take advantage of the constraint
that the ROC curves are common across~$X$.

The study motivating this research is a cross-sectional study conducted
by the Early Detection Research Network assessing a urine biomarker for
prostate cancer, Prostate Cancer Antigen 3 (PCA3). PCA3 is a
prostate-specific noncoding mRNA overexpressed in prostate tumors
[\citet
{DerasEtal2008}]. Among 576 men who were biopsied due to an elevated
prostate specific antigen, half had had a previous (negative) biopsy
and half had not had a previous biopsy. Researchers were interested in
assessing the risk prediction and classification capacity of PCA3 in
both populations, the initial biopsy population and the repeat biopsy
population. Interestingly, it was observed that the empirical ROC
curves for PCA3 were very similar in the two populations [Figure \ref{fig1}(b)],
even if it is well known that the initial biopsy population has a
higher prevalence of prostate cancer. This scenario raises the question
of how to combine data from the two populations in such a way as to
incorporate this common ROC condition when evaluating PCA3. In this
example, $X=0$ indicates the initial biopsy population and $X=1$
indicates the repeat biopsy population.

In general, let $X$ denote the population and suppose that in
preliminary analyses a test for equality of ROC curves across
populations has been accepted. The comparison between ROC curves in two
populations can be made through the comparison between distributions of
case placement values [\citet{HuangAndPepe2009a}]. For example,
comparison of the area under the ROC curve (AUC) between populations
corresponds to comparison of mean case placement values. Moreover,
comparison can be made with respect to the Wilcoxon Rank Sum statistic
of case placement values [\citet{HuangAndPepe2009a}]. An alternative
test may be based on fitting a model of the form (\ref{eqnRiskZ}) and
using a Wald test to determine if terms involving $X$ in $G(U,X,\beta)$
can be eliminated. We next propose methods to estimate the risk model,
points on the ROC curve, $\ROC_x(t)$, and the inverse of points on the
predictiveness curve, $\operatorname{CDF}_{Rx}(p)$, where the predictiveness curve is
defined as the quantile curve of the disease risk [\citet
{BuraAndGastwirth2001,HuangEtal2007,PepeEtal2008}]. In other words,
the inverse of points on the predictiveness curve is the distribution
of the disease risk, $\operatorname{CDF}_{Rx}(p)=P\{\operatorname{Risk}(X,Y)\le p|X=x\}$, where
$\operatorname{Risk}(X,Y)=P(D=1|X,Y)$. Without loss of generality, assume $G(U,X,\beta
)$ in (\ref{eqnRiskZ}) can be modeled with $\beta_0+\beta_1^Tr(U)$, a
monotone decreasing function of $U$, such that
%
\begin{equation}
\logit P(D=1|U,X)=\logit P(D=1|X)+ \beta_0+\beta_1^Tr(U).
\label{eqnRiskZZ}
\end{equation}
Different prevalences and the common ROC curve are naturally and
explicitly modeled in (\ref{eqnRiskZZ}), with the ROC curve entirely
determined by the coefficients of the risk model. None of the existing
methods we are aware of can do so. In particular, consider the standard
method of fitting logistic regression to the marker $Y$ on its original
scale and using the logit score to generate the ROC curve in each
population. Since the corresponding ROC curve depends on both the risk
model coefficients and the marker\ distribution in each population, the
common ROC assumption cannot be incorporated into the logistic risk
model without specifying the marker distributions.

\subsection{Estimated empirical likelihood estimators}\label{sec3.2}

First, we consider an empirical likelihood method. The key idea is to
maximize the estimated empirical likelihood of the observed $\hat{U}$
in the sample. The empirical likelihood method has previously been used
for evaluating goodness of fit for logistic regression models and for
constructing ROC curves based on raw marker measures only [\citet
{QinAndLawless1994}, Qin and Zhang (\citeyear{QinAndZhang1997,QinAndZhang2003})]. Here we extend
it to accommodate additional covariates using standardized marker
values that are estimated.

Let $F_{Dx}^U$ and $F_{\bar{D}x}^U$ denote the CDFs of $U$ in the case
and control populations, respectively, with covariate $X=x$. The
logistic regression model (\ref{eqnRiskZZ}) implies an exponential tilt
relationship between the probability densities of $U$ among cases and
controls: $dF_{Dx}^U(u)=\exp\{G(u,x,\beta)\}\,dF^U_{\bar{D}x}(u)$. Let
$A_{Dx}$ and $A_{\bar{D}x}$ be the set of cases and controls in the
sample with $X=x$ and let $n_{Dx}$ and $n_{\bar{D}x}$ be the
corresponding sample sizes. Let $A_D=\bigcup_xA_{Dx}$ and $A_{\bar
{D}}=\bigcup_xA_{\bar{D}x}$, let $n_{{D}}$ and $n_{\bar{D}}$ be the
total number of cases and controls, and let $n=n_D+n_{\bar{D}}$. With
$i$ indicating study subject, suppose we know the true $U_i, i=1,\ldots,n$, the empirical likelihood given ${U}_i$ is
\begin{eqnarray}
\label{eqnEMPP}&& {\cal{L}} \bigl(\beta_0,\beta_1,F^U_{\bar{D}x}
\bigr)\nonumber\\
&&\qquad=\prod_{x}\prod
_{i\in
A_{\bar
{D}x}}\,dF^U_{\bar{D}x}({U}_{i})\prod
_{x}\prod_{i\in
A_{{D}x}}\,dF^U_{{D}x}({U}_{i})
\\
&&\qquad=\prod_{x}\prod
_{i\in A_{\bar{D}x}}\,dF^U_{\bar{D}x}({U}_{i})\prod
_{x}\prod_{i\in A_{{D}x}}\exp
\bigl\{\beta_0+\beta_1^T r({U}_i)
\bigr\} \,dF^U_{\bar{D}x}({U}_{i}).\nonumber
\end{eqnarray}
By definition, for controls the conditional distribution of $U$ given
$X$ is $\operatorname{uniform} (0,1)$. Moreover, for cases the conditional distribution
of $U$ given $X$ is the common ROC curve. Therefore, the distributions
of ${U}$ conditional on disease status are the same across populations.
Henceforth, we let $p_i$ and $\exp \{\beta_0+\beta_1^T
r({U}_{i}) \}p_i$ be the ``common'' density of ${U}$ for a control
or a case in the sample and the empirical likelihood (\ref{eqnEMPP}) becomes
%
\begin{eqnarray}\label{eqnComb}
&&\Biggl\{\prod_{i=1}^n p_i
\Biggr\} \biggl[\prod_{x}\prod
_{i\in{A_{Dx}}}\exp \bigl\{\beta_0+\beta_1^T
r({U}_{i}) \bigr\} \biggr]
\nonumber
\\[-8pt]
\\[-8pt]
\nonumber
&&\qquad = \Biggl\{\prod
_{i=1}^n p_i \Biggr\} \biggl[\prod
_{i\in A_D}\exp \bigl\{\beta _0+
\beta_1^T r({U}_{i}) \bigr\} \biggr],
\end{eqnarray}
subject to $\sum_{i=1}^n p_i=1$ and $\sum_{i=1}^n \exp \{\beta
_0+\beta_1^T r({U}_i) \}p_i=1$.

This empirical likelihood can be maximized using a Lagrange multiplier
approach by solving the equation
\begin{eqnarray*}
&&\sum_{i=1}^n \operatorname{log}(p_i)+\sum
_{i\in A_D} \bigl\{\beta_0+
\beta_1^T r({U}_{i}) \bigr\} -
\lambda_1\sum_{i=1}^n
(p_i-1)\\
&&\qquad{}-\lambda_2\sum_{i=1}^n
\bigl[\exp \bigl\{ \beta_0+\beta_1^T
r({U}_i) \bigr\}p_i-1 \bigr]=0.
\end{eqnarray*}
Consequently, $(\hat{\beta}_0$, $\hat{\beta}_1)$, the maximum
likelihood estimates of $(\beta_0$, $\beta_1)$, satisfy the following
system of score equations:
\begin{eqnarray*}
\frac{\partial l(\beta_0,\beta_1)}{\partial\beta_0}&=&n_D-\sum_{i=1}^n
\frac{({n_D}/{n_{\bar{D}}})\exp \{\beta_0+\beta
_1^Tr({U}_i) \}}{1+({n_D}/{n_{\bar{D}}})\exp \{\beta
_0+\beta
_1^Tr({U}_i) \}}=0,
\\
\frac{\partial l(\beta_0,\beta_1)}{\partial\beta_1}&=&\sum
_{i\in
A_D}r({U}_{i})-\sum
_{i=1}^n\frac{r({U}_i)({n_D}/{n_{\bar
{D}}})\exp
\{\beta_0+\beta_1^Tr({U}_i) \}}{1+({n_D}/{n_{\bar
{D}}})\exp
\{\beta_0+\beta_1^Tr({U}_i) \}}=0, \label{eqnScore}
\end{eqnarray*}
which are the score equations for $(\beta_0,\beta_1)$ if we apply a
prospective logistic model
$\logit \{P(D=1|{U}) \}=\beta_0+\beta_1^Tr({U})$ to the data
with offset $n_D/n_{\bar{D}}$. The maximum likelihood estimate of $p_i$
is $
\hat{p}_i=1/ [n_{\bar{D}}+n_D\exp \{\hat{\beta}_0+\hat
{\beta
}_1^Tr({U}_i) \} ]$.

In practice, we substitute $\hat{U}_i$ for $U_i$ into the empirical
likelihood (\ref{eqnEMPP}) to get an estimated empirical likelihood,
and obtain $\hat{\beta}_0$ and $\hat{\beta}_1$ with a logistic
regression model based on $\hat{U}_i$. The corresponding estimated
empirical likelihood estimators of $F^U_{\bar{D}x}$ and $F^U_{Dx}$ are
\begin{eqnarray*}
\label{eqnF} \hat{F}^U_{\bar{D}x}(u)&=&\frac{1}{n_{\bar{D}}}\sum
_{i=1}^n \frac
{I(\hat
{U}_i\le u)}{1+({n_D}/{n_{\bar{D}}} )\exp \{\hat{\beta
}_0+\hat
{\beta}_1^Tr(\hat{U}_i) \}},
\\
\hat{F}^U_{Dx}(u)&=&\frac{1}{n_{\bar{D}}}\sum
_{i=1}^n \frac{\exp
\{
\hat{\beta}_0+\hat{\beta}_1^Tr(\hat{U}_i) \}I(\hat{U}_i\le
u)}{1+({n_D}/{n_{\bar{D}}}) \exp \{\hat{\beta}_0+\hat
{\beta
}_1^Tr(\hat{U}_i) \}},
\end{eqnarray*}
which is the same across $X$ levels. Note this procedure can be applied
whether or not there are tied values of $\hat{U}_i$ in the sample.

A piecewise differentiable and concave estimator of the common ROC
curve $\widehat{\ROC}_x(t)=\widehat{\ROC}(t)=\hat{F}_{Dx}^U{\hat
{F}_{\bar{Dx}}^{U-1}}(t)$ can be constructed based on $\hat{F}^U_{Dx}$
and $\hat{F}^U_{\bar{D}x}$ using methods analogous to those proposed by
\citet{QinAndZhang2003}, \citet{Huang2007} and \citet{HuangAndPepe2009c}. The procedure is
outlined in Appendix C of the supplementary material [\citet
{HuangEtal2013}]. Finally, we estimate $F^U_x$, the distribution of $U$
conditional on covariate $X=x$ with
$\hat{F}^U_x(u)=\hat{P}(D=0|X=x)\hat{F}^U_{\bar{D}x}(u)+\hat
{P}(D=1|X=x)\hat{F}^U_{Dx}(u)$, and estimate disease risk conditional
on $\hat{U}$ and $X$ by substituting $\hat{\beta}_0$ and $\hat
{\beta
}_1$ into (\ref{eqnRiskZZ}). Then for a particular risk threshold $p$,
we estimate $\operatorname{CDF}_{Rx}(p)$, the risk distribution at $X=x$, with $1-\hat
{F}^U_x\{u(p,x)\}$, the proportion of subjects with covariate $x$ that
have $\hat{U}$ larger than $u(p,x)=\sup_{i\in\{1,\dots,n\}}\{{\hat
{U}_i}\dvtx \hat{P}(D=1|\hat{U},X=x)\ge p\}$. We call estimators obtained
using the approach in this section the ``estimated empirical likelihood
estimators'' (EML).

\subsection{Constrained estimated maximum likelihood estimators}\label{sec3.3}

The estimat\-ed empirical likelihood method proposed in Section \ref{sec3.2} is
easy to implement using standard statistical software. In this method,
the relationship between the ROC curve and the logistic regression
model presented in Theorem \ref{th1} is utilized for combining $\hat{U}$ across
different covariate levels to estimate a common distribution as in
(\ref
{eqnComb}), under a common ROC assumption. Based on a discrete support
for~$\hat{U}$, the corresponding ROC curve estimate is piecewise
differentiable. As we will show next, an alternative way to exploit the
relationship in Theorem \ref{th1} is to use it as a constraint directly when
estimating parameters in the risk model. Since the ROC curve is
completely specified by the coefficients in the logistic regression
model~(\ref{eqnRiskZZ}), this procedure leads to a smooth ROC curve estimate.


Observe that $G(t,x,\beta)=\log\ROC_x^{\prime}(t)$ implies that the
common $\ROC(t)$ is equal to
$\int_0^t\exp\{G(u,x,{\beta})\}\,du$, which is independent of
$x$ and can be estimated by replacing $\beta$ with a consistent
estimate based on the logistic regression model~(\ref{eqnRiskZZ}).
Unlike the typical logistic regression model where there is no
constraint on the parameter space, however, the risk model~(\ref
{eqnRiskZZ}) based on the standardized marker has an implicit
constraint: $\operatorname{ROC}(1)=\int_0^1\exp\{\beta_0+\beta_1^Tr(t)\}\,dt=1$ by
definition of the ROC curve. The standard log-likelihood for $D$
conditional on $U$ and $X$ based on the logistic model is
%
\begin{equation}
l=\sum_{i=1}^n D_i\log \bigl
\{P(D_i=1|{U}_i,X_i) \bigr
\}+(1-D_i)\log \bigl\{ 1-P(D_i=1|{U}_i,X_i)
\bigr\}. \label{eqnLik}\hspace*{-35pt}
\end{equation}
We propose to maximize the estimated version of this log-likelihood
(\ref{eqnLik}) by substituting $\hat{U}$ for $U$, with the additional
constraint that
$\int_0^1\exp\{\beta_0+\beta_1^Tr(t)\}\,dt=1$.

The method used to enforce this constraint in the estimation procedure
depends on the complexity of $\beta_1$. For example, for a univariate
$\beta_1$, $\beta_0$ can be represented by a closed-form function of
$\beta_1\dvtx \beta_0=\log [\beta_1/\{\exp(\beta_1)-1\}
]$ as
$\ROC(1)=\{\exp(\beta_0+\beta_1)-\exp(\beta_0)\}/\beta_1=1$. For more
complicated models, numerical methods are needed to represent $\beta_0$
as a function of $\beta_1$. Let $(\hat{\beta}_0,\hat{\beta}_1)$ be the
estimate of $\beta$ that maximizes the constrained estimated maximum
likelihood. We estimate the common ROC curve with $\widehat{\ROC
}_x(t)=\widehat{\ROC}(t)=\int_0^t\exp\{\hat{\beta}_0+\hat{\beta
}_1^Tr(u)\}\,du$.

The CDF of risk conditional on $X=x$ can be derived from $\widehat
{\ROC
}_x$ and the disease prevalence estimate $\hat{P}(D=1|X=x)$ by
exploiting the relationship between the ROC curve and the risk
distribution shown in \citet{HuangAndPepe2009b}. Specifically, for
$p\in(0,1)$, $\operatorname{CDF}_{Rx}(p)$ can be estimated by $1-\{1-\hat
{P}(D=1|X=x)\}t-\hat{P}(D=1|X=x) \widehat{\ROC}_x(t)$, where $t$
satisfies
\begin{eqnarray*}
&&\hat{P}(D=1|X=x) \widehat{\ROC}{}^{\prime}_x(t)/ \bigl\{
\hat{P}(D=1|X=x) \widehat{\ROC}{}^{\prime}_x(t)+1-\hat{P}(D=1|X=x)
\bigr\}\\
&&\qquad=p.
\end{eqnarray*}
We call estimators obtained using the methods in this section the
``constrained estimated maximum likelihood estimators'' (CML).

\subsection{Connection to and modification of existing methods}\label{sec3.4}

Our approach to fitting a common ROC curve across populations is
similar in spirit to the covariate-adjusted ROC curve proposed by
Janes and Pepe (\citeyear
{JanesAndPepe2008,JanesAndPepe2009}), which is defined as a weighted
average of covariate-specific ROC curves
$
{\cal A}\ROC(t)=\int\ROC_x(t)\,dF_D(x),
$
where $F_D(x)$ is the CDF of $X$ among diseased case subjects. When
$\ROC_x$ is the same for different values of $x$, ${\cal{A}}\ROC$ is
the common ROC curve. Janes and Pepe
(\citeyear{JanesAndPepe2008,JanesAndPepe2009})
proposed estimating ${\cal{A}}\ROC$ nonparametrically using the
empirical CDF of $\hat{U}$ for all cases, where $\hat{U}_i=\sum_{j=1}^{n_{\bar{D}X_i}}I(Y_i> Y_j)/n_{\bar{D}X_i}$, exploiting the fact
that the distribution of $U$ for diseased observations is equal to the
ROC curve. Alternatively, semiparametric methods can be used to model
the distribution of $U$, that is, the common ROC curve. One choice of
semiparametric estimator is the pseudo-likelihood estimator (PSL),
originally proposed by \citet{PepeAndCai2004} for estimating the ROC
curve in a single population. In \citet{PepeAndCai2004}, the PSL method
imposes a parametric functional form on the ROC curve such as a
binormal form, and maximizes the likelihood of $\hat{U}$ among diseased
case subjects. This approach is easy to implement and has been shown to
have good efficiency compared to other semiparametric ROC modeling
approaches. Here we modify the PSL method by modeling the derivative of
the ROC curve directly, thereby accommodating the model form implied by
the logistic regression framework (\ref{eqnRiskZZ}). Specifically, we maximize
\[
l=\sum_{i\in A_D}\log \,dF_{Dx}^U(
\hat{U}_{i})=\sum_{i\in A_D}\log \ROC
_x^{\prime}(\hat{U}_{i})=\sum
_{i\in A_D} \bigl\{\beta_0+\beta _1^Tr(
\hat {U}_{i}) \bigr\}.
\]
In addition, we enforce the constraint $\ROC(1)=1$ as in the CML
method. Since the PSL method uses only the standardized marker value in
diseased observations, we expect that some efficiency could be gained
from our method by including standardized marker values for nondiseased
control observations as well.

\section{Application to simulated data}\label{sec4}

To mimic the setting in the PCA3 example, we simulated random samples
from two populations with disease prevalences equal to 0.44 and 0.27,
respectively. Marker distributions for controls are $N(0,1)$ in
population 1 and $N(1,1)$ in population 2. Marker distributions for cases
from each population are chosen to achieve a common ROC curve with
$\ROC
^{\prime}(t)=\exp(\beta_0+\beta_1t+\beta_2t^2)$. With $X$ being the
population indicator variable, $X=1$ for population 2, the risk model is
%
\begin{equation}
\logit P(D=1|U,X)= \alpha_0 + \alpha_1X +
\beta_0+\beta_1U+\beta _2U^2,
\label{eqnRiskSS}
\end{equation}
where the offset terms $\alpha_0$ and $\alpha_0 + \alpha_1$ are log
odds of the prevalences in the two populations, which will be estimated
empirically from the sample and entered into the model before
estimation of the other parameters in the logistic regression model.

We generated 300 observations from each population (cf. Section \ref{sec5}: the
PCA3 data set has 267 men in each population). We compared the CML,
EML, and PSL estimators for estimating: (i) $\beta=(\beta_0,\beta
_1,\beta_2)$; (ii) $\ROC(t)$ for $t=0.1,0.3,0.5,0.7,0.9$; (iii) the
risk distribution, $\operatorname{CDF}_{Rx}(p)$ versus $p$ in each population for $p$
equal to 10\%, 30\%, 50\%, 70\%, and 90\%; and (iv)
$\operatorname{Risk}(y|x)=P(D=1|Y=y,X=x)$ in each population, for $y$ corresponding to
$p$ in (iii). In each simulation, the standardized marker value $U$'s
are estimated based on nonparametric CDF estimates of the control
distribution in each population. For each estimator, we used both the
population-specific approach where only samples from the target
population are used for estimation and the combined-data approach where
we estimate a common ROC curve across populations.
We performed 10,000 Monte Carlo simulations with 500 bootstrap samples
for each simulated data set to construct confidence intervals

All three estimators, EML, CML, and PSL, have negligible bias (Table~\ref{tab1}).
Moreover, coverage of their 95\% bootstrap percentile
confidence intervals are all close to the nominal level (Table \ref
{tab3}). We note that for a sample of size 300 in each population, an
alternative to percentile bootstrap confidence intervals are Wald
confidence intervals with a bootstrap standard error estimate, which
can achieve reasonable coverage with a smaller bootstrap size such as
50; yet for smaller sample sizes (e.g., 100 in each population), they
tend to have an under-coverage problem (details omitted). We recommend
percentile bootstrap confidence intervals for our estimators in
general, as they have better performance overall.

\begin{sidewaystable}
\tablewidth=\textwidth
\tabcolsep=0pt
\caption{Bias of different estimators (multiplied by 1000 and rounded) using
data from the target population or based on the combined-data analysis.
EML is the estimated empirical likelihood estimator, CML is the
constrained estimated maximum likelihood estimator, PSL is the
constrained pseudolikelihood estimator. Standard errors of biases are
less than 0.03 for $\beta$ estimates, less than 0.001 for estimates of
$\operatorname{ROC}(t)$ and $\operatorname{Risk}(y|x)$, and less than 0.002 for estimates of
$\operatorname{CDF}_{Rx}(p)$}\label{tab1}
%
{\fontsize{7.5}{9.5}\selectfont{
\begin{tabular*}{\textwidth}{@{\extracolsep{4in minus 4in}}ld{2.2}d{2.1}d{2.1}d{2.1}d{3.1}d{3.1}d{2.1}d{2.2}d{2.1}d{2.1}d{2.1}d{3.1}d{3.1}d{2.1}@{}}
\hline
&& \multicolumn{6}{c}{\textbf{Population 1}} && \multicolumn{6}{c}{\textbf{Population
2}}\\[-6pt]
&& \multicolumn{6}{c}{\hrulefill} && \multicolumn{6}{c@{}}{\hrulefill}\\
&& \multicolumn{3}{c}{\textbf{Population-specific}} & \multicolumn
{3}{c}{\textbf{Combined-data analysis}} &&\multicolumn
{3}{c}{\textbf{Population-specific}} & \multicolumn{3}{c}{\textbf{Combined-data
analysis}}\\[-6pt]
&& \multicolumn{3}{c}{\hrulefill} & \multicolumn
{3}{c}{\hrulefill} &&\multicolumn
{3}{c}{\hrulefill} & \multicolumn{3}{c@{}}{\hrulefill}\\
&& \multicolumn{1}{c}{\textbf{EML}} &
\multicolumn{1}{c}{\textbf{CML}} & \multicolumn{1}{c}{\textbf{PSL}} & \multicolumn{1}{c}{\textbf{EML}} & \multicolumn{1}{c}{\textbf{CML}} &
\multicolumn{1}{c}{\textbf{PSL}} && \multicolumn{1}{c}{\textbf{EML}} &
\multicolumn{1}{c}{\textbf{CML}} & \multicolumn{1}{c}{\textbf{PSL}} & \multicolumn{1}{c}{\textbf{EML}} &
\multicolumn{1}{c}{\textbf{CML}} & \multicolumn{1}{c@{}}{\textbf{PSL}}\\[-6pt]
&& \multicolumn{6}{c}{\hrulefill} && \multicolumn{6}{c@{}}{\hrulefill}\\
&\multicolumn{1}{c}{\textbf{True}} & \multicolumn{6}{c}{\textbf{Bias}$\bolds{\times1000}$} &\multicolumn{1}{c}{\textbf{True}} & \multicolumn
{6}{c@{}}{\textbf{Bias}$\bolds{\times1000}$}\\
\hline
&& \multicolumn{13}{c}{$\beta$}\\
$\beta_0$ &-4 &13 & 2 & -6 & 22 & 13 & 3 &-4 & 10 & 2 & -8
& 22 & 13 & 3 \\
$\beta_1$ &1 & -73 & -52 & -30 & -120 & -107 & -72 &1 & -54 & -41 &
0 & -120 & -107 & -72 \\
$\beta_2$ &2 &14 & -1 & 5 & 85 & 75 & 66 &2 & -36 & -45 & -65 & 85
& 75 & 66 \\[6pt]
&& \multicolumn{13}{c}{$\ROC(t)$}\\
$t=0.1$ &0.27 & 5.9 & 4.2 & 2.5 & 5.9 & 4.5 & 2.5 &0.27& 5.8 & 5.1 &
3.0 & 5.9 & 4.5 & 2.5 \\
$t=0.3$ &0.59 &3.6 & 2.8 & -0.1 & 4.1 & 3.8 & 0.6 &0.59 & 4.3 & 4.0
& 1.2 & 4.1 & 3.8 & 0.6 \\
$t=0.5$ &0.77 & 1.6 & 1.5 & -1.0 & 1.6 & 1.8 & -0.7 &0.77 & 2.6 &
2.7 & 0.6 & 1.6 & 1.8 & -0.7 \\
$t=0.7$ &0.89 &-0.1 & 0.0 & -1.6 & -0.2 & 0.0 & -1.6 & 0.89& 0.3 &
0.5 & -0.8 & -0.2 & 0.0 & -1.6 \\
$t=0.9$ & 0.97 & -1.0 & -1.0 & -1.6 & -0.8 & -0.7 & -1.3 &0.97 &
-1.2 & -1.2 & -1.6 & -0.8 & -0.7 & -1.3 \\[6pt]
&& \multicolumn{13}{c}{$\operatorname{CDF}_{Rx}(p)$}\\
$p=(0.22, 0.12)^{\star}$ & 0.10 & 7.3 & 7.6 & 3.8 & 0.8 & 1.4 & -3.3
&0.10 & 22.2 & 22.8 & 18.9 & 10.4 & 10.9 & 5.9 \\
$p=(0.30, 0.16)$ & 0.30 &-1.2 & -1.7 & -4.5 & 3.1 & 3.0 & -0.1 &0.30 &
-10.6 & -10.3 & -15.0 & -1.7 & -1.6 & -6.4 \\
$p=(0.43, 0.23)$ & 0.50 &-1.0 & -2.4 & -2.1 & 2.3 & 1.2 & 1.3 &0.50& 0.6
& -0.3 & -1.2 & 4.5 & 3.4 & 2.6 \\
$p=(0.56, 0.35)$ & 0.70&4.0 & 3.4 & 5.9 & 2.2 & 1.3 & 3.8 &0.70 & 0.9 &
0.4 & 1.3 & 2.5 & 1.3 & 2.7 \\
$p=(0.67, 0.48)$ & 0.90&-4.2 & -1.7 & 1.3 & -4.7 & -3.1 & 0.7 &0.90 &
0.4 & 1.2 & 3.7 & -1.0 & 0.1 & 3.1 \\[6pt]
&& \multicolumn{13}{c}{$\operatorname{Risk}(y|x)$}\\
$y=(-1.1, -0.2)^{\star}$ &0.22& -1.3 & -2.0 & 0.4 & -0.2 & -1.1 & 1.3
&0.12& -0.6 & -1.0 & 0.2 & 0.5 & 0.0 & 1.5 \\
$y=(-0.24, -0.64)$ &0.31&-2.2 & -3.0 & -1.5 & -2.2 & -3.2 & -1.4 &0.16 &
-2.6 & -3.1 & -2.0 & -1.4 & -2.0 & -0.8 \\
$y=(0.33, 1.20)$ &0.43& 0.1 & -1.3 & -1.2 & -1.0 & -2.4 & -1.8 &0.24 &
-0.7 & -1.5 & -0.7 & -1.1 & -2.0 & -1.4 \\
$y=(0.88, 1.75)$ &0.56&3.0 & 1.1 & 0.5 & 1.7 & -0.3 & -0.5&0.35 & 2.6 &
1.2 & 1.0 & 0.9 & -0.3 & -1.0 \\
$y=(1.6, 2.5)$ &0.67&1.7 & -0.5 & -1.4 & 2.7 & 0.8 & -0.4 &0.48& 3.2 &
1.4 & 0.2 & 3.4 & 1.6 & -0.2 \\
\hline
\end{tabular*}}}
\tabnotetext[]{}{$^{\star}$: the values separated by commas correspond to population 1
and population 2, respectively.}
\end{sidewaystable}
%

\begin{sidewaystable}
\tablewidth=\textwidth
\caption{Coverage of 95\% percentile bootstrap confidence intervals (subtracted
by 95.0 and then multiplied by 10), based on the estimated empirical
likelihood estimator (EML), the constrained estimated maximum
likelihood estimator (CML), and the pseudo likelihood estimator (PSL)
using data only from the target population or based on the
combined-data analysis. Standard errors of entries do not exceed 0.4\%
of the value}\label{tab3}
{\fontsize{8.5}{10.5}\selectfont{
\begin{tabular*}{\textwidth}{@{\extracolsep{\fill}}ld{2.0}d{2.0}d{2.0}d{2.0}d{2.0}d{2.0}d{2.0}d{2.0}d{2.0}d{2.0}d{2.0}d{2.0}@{}}
\hline
& \multicolumn{6}{c}{\textbf{Population 1}} & \multicolumn{6}{c}{\textbf{Population
2}}\\[-6pt]
& \multicolumn{6}{c}{\hrulefill} & \multicolumn{6}{c@{}}{\hrulefill}\\
& \multicolumn{3}{c}{\textbf{Population-specific}} & \multicolumn
{3}{c}{\textbf{Combined-data}} &\multicolumn{3}{c}{\textbf{Population-specific}} &
\multicolumn{3}{c@{}}{\textbf{Combined-data}}\\[-6pt]
& \multicolumn{3}{c}{\hrulefill} & \multicolumn
{3}{c}{\hrulefill} &\multicolumn{3}{c}{\hrulefill} &
\multicolumn{3}{c@{}}{\hrulefill}\\
& \multicolumn{1}{c}{\textbf{EML}} & \multicolumn{1}{c}{\textbf{CML}} &
\multicolumn{1}{c}{\textbf{PSL}} & \multicolumn{1}{c}{\textbf{EML}} & \multicolumn{1}{c}{\textbf{CML}}& \multicolumn{1}{c}{\textbf{PSL}} &
\multicolumn{1}{c}{\textbf{EML}} & \multicolumn{1}{c}{\textbf{CML}}& \multicolumn{1}{c}{\textbf{PSL}} & \multicolumn{1}{c}{\textbf{EML}} & \multicolumn{1}{c}{\textbf{CML}} &
\multicolumn{1}{c@{}}{\textbf{PSL}}\\
\hline
& \multicolumn{11}{c}{$\beta$}\\
$\beta_0=-3.8$ & -10 & -6 & -4 & -17 & -8 & -2 & -9 & -8 & -3 & -17 &
-8 & -2 \\
$\beta_1=1.2$ & -5 & -5 & -8 & -8 & -6 & 0 & -7 & -8 & -2 & -8 & -6 & 0
\\
$\beta_2=1.5$ & -5 & -4 & -5 & -5 & -1 & -2 & -7 & -7 & -4 & -5 & -1 &
-2 \\[6pt]
& \multicolumn{11}{c}{$\ROC(t)$}\\
$t=0.1$ & -7 & -5 & -3 & -16 & -10 & -1 & -11 & -6 & -5 & -16 & -10 & -1
\\
$t=0.3$ & -8 & -8 & -4 & -13 & -14 & -10 & -9 & -8 & -6 & -13 & -14 &
-10 \\
$t=0.5$ & -8 & -7 & -2 & -10 & -11 & -6 & -14 & -13 & -9 & -10 & -11 &
-6 \\
$t=0.7$ & -7 & -8 & -3 & -5 & -6 & -4 & -12 & -15 & -9 & -5 & -6 & -4 \\
$t=0.9$ & -5 & -6 & -4 & -5 & -6 & -5 & -13 & -14 & -11 & -5 & -6 & -5
\\[3pt]
& \multicolumn{11}{c}{$\operatorname{CDF}_{Rx}(p)$}\\
$p=(0.22, 0.12)^{\star}$ & -2 & -4 & 1 & -1 & -1 & 5 & -20 & -21 & -14 &
-4 & -8 & -1 \\
$p=(0.30, 0.16)$ & 2 & 1 & 1 & 0 & -3 & 2 & -4 & -7 & -5 & 3 & 1 & 5 \\
$p=(0.43, 0.23)$ & -1 & -3 & -4 & -2 & -3 & -3 & 8 & 5 & 6 & 1 & 4 & 2 \\
$p=(0.56, 0.35)$ & -1 & -6 & -4 & -3 & -8 & -7 & 8 & 5 & 4 & 2 & 0 & -2 \\
$p=(0.67, 0.48)$ & -11 & -8 & -6 & -7 & -7 & -3 & -5 & -6 & -3 & -5 & -4
& -4 \\[3pt]
& \multicolumn{11}{c}{$\operatorname{Risk}(y|x)$}\\
$y=(-1.1, -0.17)^{\star}$ & -4 & -6 & 0 & 0 & -1 & 2 & -19 & -21 & -15 &
-4 & -7 & -5 \\
$y=(-0.24, 0.64)$ & -2 & -5 & -3 & -2 & -7 & 0 & -10 & -12 & -6 & 3 & 0 &
5 \\
$y=(0.33, 1.2)$ & -5 & -4 & -2 & 1 & -2 & 1 & 1 & 1 & 0 & 0 & 0 & 1 \\
$y=(0.88, 1.7)$ & -5 & -1 & -1 & -2 & 0 & -1 & 5 & 6 & 6 & 4 & 3 & 3 \\
$y=(1.6, 2.5)$ & -16 & -12 & -10 & -14 & -8 & -5 & -6 & -3 & -1 & -2 &
0 & 2 \\
\hline
\end{tabular*}}}
\tabnotetext[]{}{$^{\star}$: the values separated by commas correspond to population 1
and population 2, respectively.}
\end{sidewaystable}

\begin{table}
\tabcolsep=0pt
\caption{Efficiency of other estimators (using data from the target population
or based on the combined-data analysis) relative to the constrained
maximum likelihood estimator calculated using data only from the target
population. Standard errors of entries do not exceed 3\% of the value}\label{tab2}
\begin{tabular*}{\textwidth}{@{\extracolsep{\fill}}lcccccccccc@{}}
\hline
& \multicolumn{5}{c}{\textbf{Population 1}} & \multicolumn{5}{c@{}}{\textbf{Population
2}}\\[-6pt]
& \multicolumn{5}{c}{\hrulefill} & \multicolumn{5}{c@{}}{\hrulefill}\\
& \multicolumn{2}{c}{\textbf{Pop-specific}} & \multicolumn{3}{c}{\textbf{Combined-data}}
&\multicolumn{2}{c}{\textbf{Pop-specific}} & \multicolumn{3}{c@{}}{\textbf{Combined-data}}\\[-6pt]
& \multicolumn{2}{c}{\hrulefill} & \multicolumn{3}{c}{\hrulefill}
&\multicolumn{2}{c}{\hrulefill} & \multicolumn
{3}{c@{}}{\hrulefill}\\
& \textbf{EML} & \textbf{PSL} & \textbf{EML} & \textbf{CML} & \textbf{PSL} & \textbf{EML} & \textbf{PSL} & \textbf{EML} & \textbf{CML} & \textbf{PSL}\\
\hline
& \multicolumn{9}{c}{$\beta$}\\
$\beta_0=-3.8$ & 0.98 & 0.94 & 1.81 & 1.89 & 1.77 & 0.98 & 1.00 & 2.24
& 2.34 & 2.20 \\
$\beta_1=1.2$ & 0.99 & 0.92 & 1.81 & 1.87 & 1.73 & 0.99 & 0.99 & 2.27 &
2.35 & 2.18 \\
$\beta_2=1.5$ & 0.99 & 0.91 & 1.80 & 1.85 & 1.70 & 0.99 & 0.98 & 2.35 &
2.42 & 2.23 \\[3pt]
& \multicolumn{9}{c}{$\ROC(t)$}\\
$t=0.1$ & 0.96 & 0.95 & 1.75 & 1.84 & 1.76 & 0.99 & 1.01 & 2.16 & 2.28 &
2.18 \\
$t=0.3$ & 0.99 & 0.98 & 1.77 & 1.81 & 1.77 & 1.00 & 1.01 & 2.25 & 2.31 &
2.25 \\
$t=0.5$ & 0.99 & 0.98 & 1.71 & 1.75 & 1.71 & 0.99 & 0.99 & 2.33 & 2.38 &
2.33 \\
$t=0.7$ & 0.98 & 0.95 & 1.66 & 1.71 & 1.62 & 0.98 & 0.97 & 2.33 & 2.39 &
2.27 \\
$t=0.9$ & 0.97 & 0.92 & 1.66 & 1.72 & 1.59 & 0.98 & 0.96 & 2.31 & 2.39 &
2.21 \\[3pt]
& \multicolumn{9}{c}{$\operatorname{CDF}_{Rx}(p)$}\\
$p=(0.22, 0.12)^{\star}$ & 0.99 & 1.01 & 1.14 & 1.15 & 1.17 & 1.00 & 1.02
& 1.28 & 1.29 & 1.32 \\
$p=(0.30, 0.16)$ & 0.98 & 0.95 & 1.31 & 1.35 & 1.26 & 0.99 & 0.97 & 1.39
& 1.43 & 1.36 \\
$p=(0.43, 0.23)$ & 0.96 & 0.96 & 1.20 & 1.24 & 1.19 & 0.99 & 0.98 & 1.35
& 1.40 & 1.34 \\
$p=(0.56, 0.35)$ & 0.98 & 0.96 & 1.37 & 1.38 & 1.35 & 0.99 & 0.98 & 1.44
& 1.46 & 1.42 \\
$p=(0.67, 0.48)$ & 0.98 & 1.00 & 1.24 & 1.25 & 1.25 & 0.99 & 1.01 & 1.26
& 1.28 & 1.26 \\[3pt]
& \multicolumn{9}{c}{$\operatorname{Risk}(y|x)$}\\
$y=(-1.1, -1.2)^{\star}$ & 1.00 & 0.96 & 1.51 & 1.53 & 1.47 & 0.99 &
0.97 & 1.98 & 2.00 & 1.91 \\
$y=(-0.24, -0.36)$ & 0.99 & 0.95 & 1.32 & 1.32 & 1.29 & 1.00 & 1.00 &
1.51 & 1.55 & 1.49 \\
$y=(0.33, 0.20)$ & 0.99 & 0.92 & 1.16 & 1.14 & 1.13 & 0.99 & 0.98 & 1.35
& 1.40 & 1.32 \\
$y=(0.88, 0.75)$ & 1.00 & 0.99 & 1.08 & 1.08 & 1.08 & 0.99 & 1.00 & 1.10
& 1.12 & 1.10 \\
$y=(1.6, 1.5)$ & 1.00 & 0.93 & 1.40 & 1.45 & 1.34 & 0.99 & 0.98 & 1.53
& 1.53 & 1.50 \\
\hline
\end{tabular*}
\tabnotetext[]{}{$^{\star}$: the values separated by commas correspond to population 1
and population 2, respectively.}
\end{table}

Table \ref{tab2} shows the efficiencies of the three estimators
relative to each other. We use the population-specific CML estimator as
the reference in this table so the entries are the variances of the CML
estimator that employs data for the target population only relative to
variances of the estimators. In general, the EML estimator appears to
be slightly less efficient than the CML estimator, and both the CML and
EML estimators are slightly more efficient than the PSL estimator. Most
importantly, Table \ref{tab2} shows that the combined-data analysis
dramatically increases efficiency compared to using only the
population-specific data. The magnitude of the efficiency gain varies
with the measures and the target population of interest. For example,
when population 1 is the target population, the relative efficiency of
the combined-data analysis versus the population-specific method ranges
from 1.6--1.8 for ROC estimation, and ranges from 1.1--1.4 for risk
distribution estimation; when population 2 is the target population,
the relative efficiency of the combined-data analysis versus the
population-specific method ranges from 2.2--2.4 for ROC estimation, and
ranges from 1.3--1.5 for risk distribution estimation.

Since the nonparametric method is commonly used for estimating ROC and
${\cal{A}\ROC}$ curves, we compared their performances with the
proposed method based on the logistic regression framework.
In Table \ref{tabAROC} we present the nonparametric estimate of the ROC
curve within each population and the nonparametric ${\cal{A}}\ROC$
estimate that combines data from the two populations to estimate the
common ROC curve. While the nonparametric estimates have minimal bias,
it appears in our simulation setting that coverage of the bootstrap
percentile confidence interval can be remarkably below nominal levels
for points at the end of the ROC curve. Moreover, we see that the
logistic regression framework gains substantial efficiency over the
nonparametric method. The efficiency of the CML estimator relative to
the nonparametric estimator varies from 1.2 to 1.9 in Table \ref{tabAROC}.

\begin{table}
\caption{Performance of the nonparametric $\ROC(t)$ and
${\cal{A}}\ROC(t)$ estimators. Standard errors of biases of $\operatorname{ROC}(t)$ or
${\cal{A}}\ROC(t)$ are less than 0.001. Standard errors for coverage of 95\%
confidence intervals do not exceed 0.7\% of the value. Standard errors
for relative efficiency do not exceed 1\% of the value}\label{tabAROC}
\begin{tabular*}{\textwidth}{@{\extracolsep{\fill}}ld{2.3}d{2.3}d{2.3}@{}}
\hline
& \multicolumn{1}{c}{\textbf{Population 1}} & \multicolumn{1}{c}{\textbf{Population 2}} & \multicolumn{1}{c@{}}{\textbf{Combined-data}} \\
\hline
\multicolumn{4}{c}{Bias$\times1000$}\\
$t=0.1$ & 0.012 & 0.014 & 0.008 \\
$t=0.3$ & 0.005 & 0.008 & 0.002 \\
$t=0.5$ & 0.004 & 0.006 & 0.002 \\
$t=0.7$ & 0.004 & 0.006 & 0.002 \\
$t=0.9$ & 0.004 & 0.006 & 0.002 \\[3pt]
\multicolumn{4}{c}{Coverage of 95\% bootstrap percentile CI}\\
$t=0.1$ & 94.4 & 95.4 & 95.6 \\
$t=0.3$ & 95.4 & 94.6 & 95.7 \\
$t=0.5$ & 94.8 & 94.2 & 95.6 \\
$t=0.7$ & 94.0 & 92.4 & 95.3 \\
$t=0.9$ & 90.2 & 82.9 & 93.8 \\[3pt]
\multicolumn{4}{c}{Efficiency of CML relative to the nonparametric
estimator}\\
$t=0.1$ & 1.83 & 1.64 & 1.84 \\
$t=0.3$ & 1.21 & 1.23 & 1.26 \\
$t=0.5$ & 1.28 & 1.29 & 1.32 \\
$t=0.7$ & 1.22 & 1.23 & 1.25 \\
$t=0.9$ & 1.72 & 1.68 & 1.88 \\
\hline
\end{tabular*}
\end{table}

We further evaluated the proposed methods by varying the simulation
settings. Results for estimating the ROC curve, risk, and risk
distribution are presented in Appendix D of the supplementary material
[\citet{HuangEtal2013}]. We examined two additional scenarios where the
common ROC curve condition holds across two populations. The first
scenario has smaller sample sizes (100 in each populations) compared to
the primary setting (Tables 1--3 in Appendix D). The second scenario has
different control marker distributions across populations [$N(0,1)$ in
population 1 and $\operatorname{logNormal}(1,1)$ in population 2] (Tables 4--6 in
Appendix D). Smaller sample sizes lead to slightly larger bias. But
overall we observe minimal bias, good coverage, and good efficiency
gain with the combined-data analysis in each of the two scenarios. We
also examined another scenario where there is a slight difference in
the ROC curves between the two populations, with the area under the ROC
curve 5\% larger in population 2 relative to population 1. In the
presence of a small difference in ROC curves, the combined-data
approach estimating the common ROC curve has slightly larger bias
compared to the population-specific approach, but in general still
maintains a good efficiency gain in terms of an appreciable drop in the
mean squared error (Tables 7--8 in Appendix D).

\section{Application to a prostate cancer data set}\label{sec5}
We illustrate our methodology using the PCA3 example data set that
includes 576 patients [\citet{DerasEtal2008}] who underwent a diagnostic
biopsy for prostate cancer due to elevated PSA levels. Among these
subjects, 267 had a previous negative biopsy and 267 subjects had no
previous biopsy.

\begin{figure}

\includegraphics{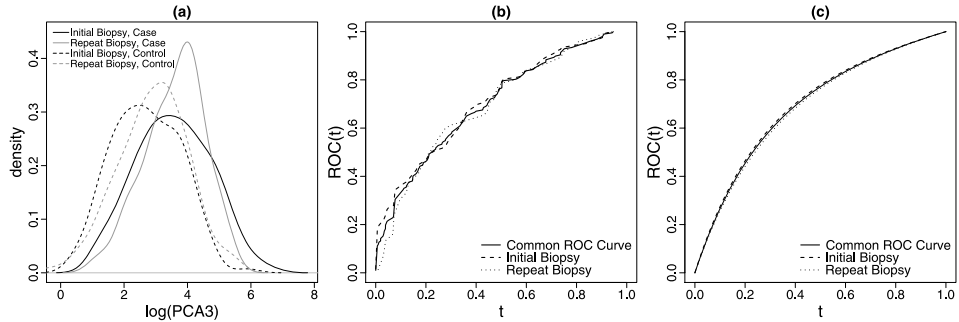}

\caption{\textup{(a)} density of $\operatorname{log}(\mathrm{PCA3})$ conditional on disease
status in
each population; \textup{(b)} nonparametric and
\textup{(c)} semiparametric ROC curves for PCA3 in initial and repeat
biopsy populations,
and the common ROC curve adjusting for population effect.}\label{fig1}
\end{figure}

Figure \ref{fig1}(a) shows probability density functions for $\operatorname{log}(\mathrm{PCA3})$
conditional on disease status, and Figure \ref{fig1}(b) shows the empirical ROC
curves in the two populations. Interestingly, although the
distributions of PCA3 conditional on disease status seem to differ
between the two populations, the two ROC curves appear similar to each
other. A test for equality of the ROC curves based on the area under
the ROC curves yields a $p$-value of 0.66. An alternative test comparing
case placement values using the Wilcoxon Rank Sum statistic yields a
$p$-value of 0.45 [\citet{HuangAndPepe2009a}]. Also presented in Figure
\ref{fig1}(b) is the nonparametric estimate of the common ROC curve [\citet
{JanesAndPepe2009}] $\widehat{{\cal{A}}\ROC}(t)=\sum_{i=1}^{n_D}I\{
Y_{Di}>\hat{S}_{\bar{D}X_i}^{-1}(t)\}/n_D$.

\begin{sidewaystable}
\tablewidth=\textwidth
\label{tab4}
\caption{The constrained maximum likelihood estimators calculated with
data from the PCA3 study}
\begin{tabular*}{\textwidth}{@{\extracolsep{\fill}}lccccc@{}}
\hline
&& \multicolumn{2}{c}{\textbf{Initial biopsy}} & \multicolumn{2}{c}{\textbf{Repeat
biopsy}}\\[-6pt]
&& \multicolumn{2}{c}{\hrulefill} & \multicolumn{2}{c@{}}{\hrulefill}\\
&& \multicolumn{1}{c}{\textbf{Combined-data}} & \multicolumn{1}{c}{\textbf{Population specific}} & \multicolumn{1}{c}{\textbf{Combined-data}} &
\multicolumn{1}{c@{}}{\textbf{Population
specific}} \\
\hline
$\ROC(t)$ &&& \\
$t=0.1$ & Est & 0.266 (0.209, 0.339) & 0.274 (0.205, 0.368) & 0.266
(0.209, 0.339)& 0.256 (0.166, 0.369) \\
& Eff$^{\star}$ & 1.57 & 1.00 & 2.53 & 1.00\\[3pt]
$t=0.3$ & Est & 0.591 (0.518, 0.671) & 0.600 (0.510, 0.704) & 0.591
(0.518, 0.671)& 0.578 (0.462, 0.697) \\
& Eff & 1.57 & 1.00 & 2.41 & 1.00\\[3pt]
$t=0.5$ & Est & 0.772 (0.710, 0.831) & 0.778 (0.703, 0.854) & 0.772
(0.710, 0.831) & 0.764 (0.670, 0.847)\\
& Eff & 1.64 & 1.00 & 2.21 & 1.00\\[3pt]
$t=0.7$ & Est & 0.886 (0.836, 0.927) & 0.889 (0.827, 0.939) & 0.886
(0.836, 0.927)& 0.882 (0.810, 0.940)\\
& Eff & 1.67 & 1.00 & 2.19 & 1.00\\[3pt]
$t=0.9$ & Est & 0.967 (0.943, 0.982) & 0.967 (0.937, 0.986) & 0.967
(0.943, 0.982)& 0.966 (0.930, 0.987)\\
& Eff & 1.67 & 1.00 & 2.32 & 1.00\\[6pt]
$\operatorname{CDF}_R(p)$ && \multicolumn{2}{c}{$p=0.65$} & \multicolumn
{2}{c}{$p=0.20$}\\
&Est & 0.855 (0.711, 1.000) & 0.843 (0.698, 1.000) & 0.418 (0.276,
0.564) & 0.406 (0.211, 0.586)\\
&Eff & 1.12 & 1.00 & 1.60 & 1.00 \\[6pt]
$\operatorname{Risk}(y)$ && \multicolumn{2}{c}{$y=60$} & \multicolumn{2}{c}{$y=20$}\\
&Est & 0.654 (0.567, 0.731) & 0.659 (0.569, 0.749) & 0.208 (0.147,
0.270) & 0.214 (0.135, 0.292) \\
&Eff & 1.22 & 1.00 & 1.59 & 1.00 \\
\hline
\end{tabular*}
\tabnotetext[]{}{Eff$^{\star}$: efficiency relative to the population-specific
estimator.}
\end{sidewaystable}


As we have noted earlier, we cannot enforce a common ROC condition by
fitting a logistic model with $Y$ on the usual scale. But fitting a
model with $Y$ on the $U$ scale such as $\logit P(D=1|U,X)=\logit
P(D=1|X) + \beta_0+\beta_1^Tr(U)$ automatically guarantees a common ROC
across $X$.
We adopt a logistic model $\logit P(D=1|U,X)=\logit P(D=1|X)+\beta
_{0}+\beta_{1}U+\beta_{2}U^2$. Figure \ref{fig1}(c) displays the ROC curve
estimates calculated by fitting this model separately in the two
populations and by using the combined-data analysis method. Observe
that the curves are all concave and are very similar to each other.
However, the combined-data analysis estimates are much more precise
(Table \ref{tab4}). Borrowing information across populations leads to an
efficiency gain of over 50\% compared to using the initial biopsy
sample only and over 100\% gain in efficiency compared to using the
repeat biopsy sample only. Comparing the CML estimates of the common
$\ROC(t)$ (combined-data analysis) with the nonparametric ${\cal
{A}}\ROC
(t)$ estimates [Figure \ref{fig1}(b)], the efficiency gains through modeling the
ROC derivative are 87\%, 41\%, 78\%, 51\%, and 74\%, respectively, for
$t=0.1, 0.3, 0.5, 0.7$, and $0.9$.

\begin{figure}

\includegraphics{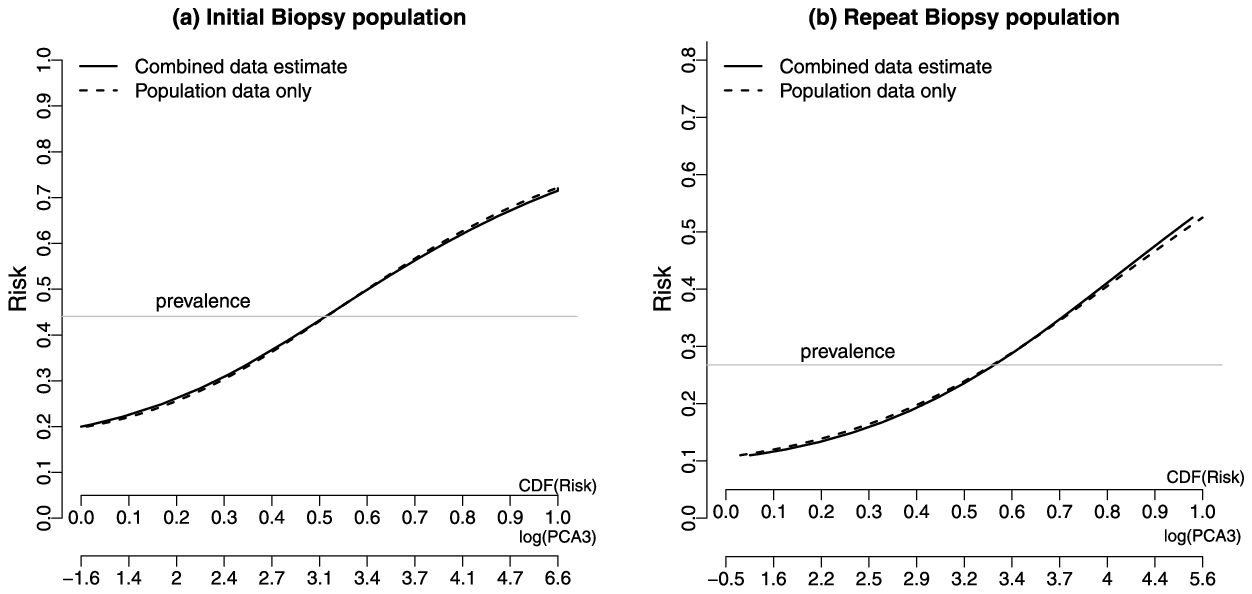}

\caption{Risk as a function of $\operatorname{log}(\mathrm{PCA3})$ and versus the CDF of Risk in
the \textup{(a)} initial biopsy and \textup{(b)} repeat biopsy populations.}
\label{fig6.1}
\end{figure}

The estimated risk is shown in Figure \ref{fig6.1} as a function of $\operatorname{log}(\mathrm{PCA3})$ and
the risk distributions in the two populations are shown as well. Again,
curves derived from fitting the risk model to each population
separately appear to be similar to those derived from the combined-data
analysis. Urologists are particularly interested in the capacity of
PCA3 to identify high risk subjects from the initial biopsy population
and low risk subjects from the repeat biopsy population. Toward this
goal, we want to have an accurate prediction of the prostate cancer
risk as a function of PCA3 among each population, and to have a good
assessment of the population impact of PCA3 in assisting treatment
decision. Here we evaluate the risk of prostate cancer at $\mathrm{PCA}3=60$,
denoted by $\operatorname{Risk}(60)$, for the former population and the risk of prostate
cancer at $\mathrm{PCA}3=20$, denoted by $\operatorname{Risk}(20)$, for the latter population. In
addition, suppose an estimated risk greater than 0.65 will lead to a
recommendation for treatment and a risk below 0.20 will lead to a
recommendation against treatment. Therefore, we assess $\operatorname{CDF}_R(0.65)$ in
the initial biopsy population and $\operatorname{CDF}_R(0.20)$ in the repeat biopsy
population. Table \ref{tab4} presents the CML estimators for those
quantities based on population-specific analysis and on combined-data
analysis. Again, both approaches result in similar estimates, but
combined-data analysis yields more precise estimates. The efficiency of
the combined-data analysis is modest for evaluating the risk prediction
capacity of PCA3 in the initial biopsy population [1.22 for estimating
$\operatorname{Risk}(60)$ and 1.12 for estimating $\operatorname{CDF}_R(0.65)$], but much larger when
the repeat biopsy population is concerned, around 1.6 for both $\operatorname{Risk}(20)$
and $\operatorname{CDF}_R(0.20)$. Our results provide useful information to urologists
regarding the value of PCA3 in treatment decision making. In
particular, based on a high risk threshold of 0.65 and a low risk
threshold of 0.20, use of PCA3 will recommend 86\% of subjects for
treatment in initial biopsy population and will spare 58\% of subjects
from treatment in repeat biopsy population.

\section{Discussion}
\label{s:discuss}

In this paper we proposed a logistic regression framework for modeling
marker values after they are standardized using the distribution in the
nondiseased control population. This sort of standardization is often
used in laboratory
and clinical medicine. For example, \citet{Frischancho1990} provides
weight and height of children
standardized relative to a healthy population of children of the same
age and gender.
Our framework provides a convenient way to connect risk modeling with
ROC analysis with many applications. For example, one can use it for
simply estimating the ROC curve from a single cohort or case--control
study, for evaluating covariate effects on biomarker performance, and
for combining data sources in evaluating biomarker performance through
the estimation of a common ROC curve when applicable, as presented in
this paper.

Covariate adjustment is an important issue in ROC curve evaluation.
When a biomarker's distribution depends on covariates, adjusting for
the covariate effect allows an evaluation of biomarker's classification
performance independent of the covariate level. Estimation of a common
ROC curve adjusting for a covariate that affects marker distribution in
a classification study is analogous to estimation of a common odds
ratio across covariate strata in an association study but requires
different techniques for covariate adjustment. While the covariates to
be adjusted for in odds ratio estimation are included in the regression
model, covariate-adjustment in ROC analysis is achieved in the step of
marker standardization. We developed two types of combined-data
analysis estimators based on a common ROC curve. The constrained
estimated maximum likelihood estimator is more efficient when there is
large variability in estimated standardized marker values across
populations, due, for example, to variation in the sizes of the
reference nondiseased sample across populations. The estimated
empirical likelihood estimator, on the other hand, is easier to
implement and may be preferable for complicated models. R code for
computing the two estimators is available upon request. An
easy-to-implement nonparametric method for covariate-adjustment in ROC
estimation has been proposed in Janes and Pepe (\citeyear{JanesAndPepe2008,JanesAndPepe2009}).
Our logistic regression based estimator provides a
much more efficient semiparametric alternative. An additional advantage
of our semiparametric method over the nonparametric method is the ease
of retrieving the ROC derivative, which can be used for various
purposes, including the derivation of the risk distribution [\citet
{HuangAndPepe2009b}]. Moreover, while the nonparametric method does not
model risk factors for risk prediction and operationally is more
suitable for discrete $X$, our framework has the same flexibility as a
traditional logistic regression model. Our method also provides a
useful addition to the semiparmatric ROC modeling field. Unlike other
existing semiparametric ROC regression methods [e.g., by \citet
{PepeAndCai2004,AlonzoAndPepe2002,DoddAndPepe2003}] that posit
assumptions on the functional form of the ROC curve (typically a
binormal ROC form), we fit a model to the ROC derivative directly. One
attractive property of this strategy is that one can easily build in
the constraint that the ROC curve is concave, which is a fundamental
attribute of proper ROC curves, whereas the traditional binormal ROC
model is not natural for ensuring concavity. Another attractive feature
of the logistic regression framework is that it accommodates
case--control sampling which is common in biomarker research studies
[\citet{PepeEtal2001}].

Moreover, as shown in Sections \ref{sec3.4} and \ref{sec4}, another novelty of the
logistic regression framework over existing ROC regression methods
based on marker standardization is the way the standardized marker
values are used. Our approach fits a prospective risk model based on
standardized marker values among all subjects and is more efficient
compared to the traditional ROC regression methods that build on
standardized marker value among cases only.

Our methods for combining data from different sources is flexible and
applies whether or not the components combined have the same study
design. For example, we can have one component being a case--control
study and the other component a cohort study. Frequency matching in
case--control studies can also be accommodated by adjusting for biased
sampling in estimation of the standardized marker value and in fitting
of the logistic regression model. The former can be conducted using the
fact that $U(y)=P(Y>y|D=0)=\sum_s P(Y>y|D=0,S=s)P(D=0|S=s)$ for cases
and controls frequency-matched within stratum $S$. The latter can be
conducted weighting the contribution of each observation to the
likelihood by the inverse of the sampling probability.

Finally, we want to point out that there are different ways to assess
our model calibrations in practice. The direct correspondence between
our risk model and the ROC model allows assessment of model calibration
based on ROC model checking techniques such as those proposed in \citet
{CaiAndZheng2007}. In our motivating example, goodness of fit can be
demonstrated graphically comparing the nonparametric estimate of the
covariate-adjusted ROC curve with our semiparametric estimates.
Alternatively, model checking can be conducted through general
Hosmer--Lemeshow type techniques developed for logistic regression
[\citet{HosmerAndLemeshow1980,HuangAndPepe2010b}].


\section*{Acknowledgments}
We thank the
Editor, Associate Editor, and referees for their constructive comments.

\begin{supplement}[id=suppA]
\stitle{Supplementary Appendix}
\slink[doi]{10.1214/13-AOAS634SUPP} 
\sdatatype{.pdf}
\sfilename{aoas634\_supp.pdf}
\sdescription{Supplement: Proof of Theorem \ref{th1}, a simulated example
referred to in Section~\ref{sec2.1}, steps of the construction of a concave ROC
curve based on the pseudoempirical
likelihood estimators, and additional simulation results. The
supplementary material would be provided at this location.}
\end{supplement}

%

%

\printaddresses

\end{document}